\documentclass[letterpaper]{article} 
\usepackage{aaai24}  
\usepackage{times}  
\usepackage{helvet}  
\usepackage{courier}  
\usepackage[hyphens]{url}  
\usepackage{graphicx} 
\usepackage{natbib}  
\usepackage{caption} 
\usepackage{algorithm}
\usepackage{algorithmic} 

\usepackage{newfloat}
\usepackage{listings}
\DeclareCaptionStyle{ruled}{labelfont=normalfont,labelsep=colon,strut=off} 
\floatstyle{ruled}
\newfloat{listing}{tb}{lst}{}
\floatname{listing}{Listing}
\title{Towards a Harms Taxonomy of AI Likeness Generation}
\author{
    Ben Bariach\textsuperscript{\rm 1},
    Bernie Hogan\textsuperscript{\rm 1},
    Keegan McBride\textsuperscript{\rm 1}
}
\affiliations{
    \textsuperscript{\rm 1}   
    University of Oxford\\

1 St. Giles, OX1 3JS\\
Oxford, United Kingdom\\
    {first}.{last}@oii.ox.ac.uk
}

\nocopyright

\begin{document}

\maketitle

\begin{abstract}
Generative artificial intelligence models, when trained on a sufficient number of a person's images, can replicate their identifying features in a photorealistic manner. We refer to this process as ``likeness generation''. Likeness-featuring synthetic outputs often present a person’s likeness without their control or consent, and may lead to harmful consequences. This paper explores philosophical and policy issues surrounding generated likeness. It begins by offering a conceptual framework for understanding likeness generation by examining the novel capabilities introduced by generative systems. The paper then establishes a definition of likeness by tracing its historical development in legal literature. Building on this foundation, we present a taxonomy of harms associated with generated likeness, derived from a comprehensive meta-analysis of relevant literature. This taxonomy categorises harms into seven distinct groups, unified by shared characteristics. Utilising this taxonomy, we raise various considerations that need to be addressed for the deployment of appropriate mitigations. Given the multitude of stakeholders involved in both the creation and distribution of likeness, we introduce concepts such as indexical sufficiency, a distinction between generation and distribution, and harms as having a context-specific nature. This work aims to serve industry, policymakers, and future academic researchers in their efforts to address the societal challenges posed by likeness generation.   
\end{abstract}

\renewcommand{\thefootnote}{\fnsymbol{footnote}}
\footnotetext[1]{This is a pre-print version of a work that has been submitted for possible publication. Copyright may be transferred without notice, after which this version may no longer be accessible.}
\renewcommand{\thefootnote}{\arabic{footnote}}

\section{Introduction}

Generative artificial intelligence models can use statistical data distributions to create images with lifelike fidelity \cite{nightingale2022ai}. When trained on diverse images of various people, these models often yield average or indistinct faces, blending traits from the training dataset. However, when trained on a sufficient number of images of a specific person, these models can produce new images of that person in different contexts, environments or artistic styles. The resulting outputs vary in their photo-realism and accuracy. Yet, they are often sufficiently photorealistic and accurate of a given person’s likeness that a viewer can recognise the subject in the synthetic image, sometimes even questioning the image's authenticity. The newfound ability to generate one's likeness may render the likeness outside of the subject’s control or consent. Once embedded in a publicly accessible model, users can generate images of the subject without technically needing their permission.

This paper explores the emergent phenomenon of likeness generation in image generation models. Given the novelty of consumer or prosumer capacities to render a synthetic image with a real likeness, many of the societal questions related to this practice are yet open. In this initial attempt to clarify some of those questions, we pose the research questions: what is likeness generation, what are the harms that it introduces to humans and society, and what considerations should be weighed in mitigating these harms?

To address these questions, we begin by exploring the technical development of likeness generation. We review the relevant technical literature to outline the likeness generation capabilities and features of generative models, as we highlight that they differ from earlier production methods colloquially called ``deepfakes'' \cite{widder2022limits}. We then we examine historical instances in which individuals experienced harm due to the use of their identifiable visual features. This examination aims to establish a shared definition for the term 'likeness'.

At the centre of this paper, we conduct a meta-analysis of multidisciplinary literature to produce a unified taxonomy for harms associated with generated likeness. Building upon this taxonomy, we surface four considerations that may assist future efforts to develop and deploy relevant mitigations. Given the multitude of stakeholders involved in the likeness generation and distribution supply chain, we use these considerations to also clarify the optimal level of their deployment.

\section{Likeness Generation as a Technical Process}

Generative models increasingly enable reproducing the personal identity of people. Audio diffusion models are capable of producing human voices \cite{Kong}, language models can emulate the communicative writing style of individuals \cite{Thoppilan} and video models exhibited considerable progress in generating realistic, stable renditions of talking human heads and facial expressions \cite{Stypulkowski, xu2024vasa1}. Generative models are also capable of reproducing non-inherent identifying features of individuals, such as imagery models producing art outputs in the distinctive oeuvre of known painters \cite{ZhangandHuang2023}. These developments highlight the expanding capacities of generative models to reproduce unique features of human identity across different modalities.

Among them, image generation models have demonstrated a particularly proficient capability of reproducing identifying features, in the form of generating a person’s appearance photorealistically. This development is rooted in progress in the field of natural image synthesis, a collection of machine learning tasks dedicated to producing images virtually indistinguishable from real-world images \cite{Brock}. Generative imagery models employ different technical approaches that involve trade-offs in parameters such as execution time, architecture, and other factors that influence output quality \cite{Bond-Taylor}. Earlier models, including Generative Adversarial Networks (GANs), Variational Autoencoders (VAEs), and autoregressive models have exhibited limitations in output fidelity, diversity, and performance when tasked with synthesising complex natural images \cite{Ho}.

Diffusion Models have recently emerged as the current state-of-the-art for producing photorealistic imagery, demonstrating exceptional efficiency and reliability in generating realistic, high-fidelity outputs that capture the complex distributions of natural images \cite{Rombach}. They are particularly capable of generating human faces without requiring fine-tuning on face-specific datasets \cite{Klemp}. These traits have led to the development of various diffusion-based foundation models, which were previously defined as the act of ``Train[ing] one model on a huge amount of data and adapt[ing] it to many applications'' \cite{CRFM}. Diffusion-based foundation models power various popular consumer facing systems that are capable of varying degrees of likeness generation fidelity, and are often developed by leading commercial AI labs. Some of the most popular diffusion foundation models are closed source models, confined to generating likeness that was pre-encoded into the model through its training on a specific dataset. 

Alongside closed-source models, open-source image generation foundation models also emerged. The ‘Stable Diffusion’ model, for example, is one such popular model. Recent advancements have introduced the ability to fine-tune such open-source models to incorporate additional objects after the model’s pre-training. Google's 'DreamBooth' technique allows for the personalisation of Diffusion Models by letting users upload additional image inputs to be reproduced in new visual renditions \cite{ruiz2023dreambooth}. Similarly, the online open-source image generation community has leveraged the Low-Rank Adaptation (LoRA) technique to tailor pre-trained models to be able to generate new objects. These techniques enable users to upload image samples depicting humans onto open-source models for their re-purposing towards the generation of human depictions, based on text prompts corresponding with their labels. Additionally, users can exert greater control over how such objects are depicted using the 'ControlNet' technique, which enhances the control of diffusion models in various ways, such as manipulating human poses \cite{ZhangAgra} and transfer of faces from one photo to another through IP Adapter models \cite{ye2023ip-adapter}.

Beyond their text-to-image functionality, image generation models may include an 'inpainting' capability, which generates new visual content within a masked region of an image \cite{Rombach}. It enables editing existing images by masking a specific region and using a text prompt to populate that region with the desired content, which was demonstrated as particularly efficient in photorealistic imagery \cite{Nichol2022}. The inpainting capability enables the seamless manipulation of imagery, including those featuring a person’s likeness.  

The open-source property of Stable Diffusion model has given rise to vibrant online communities, centred around its use and re-training. These communities are composed of users and developers who customise the model to incorporate new elements into it, including identifiable humans. Besides their convergence on popular online platforms such as Reddit, websites dedicated to Stable Diffusion have emerged for the sharing and downloading of customised Stable Diffusion models. 'Civitai' (https://civitai.com/), one such platform, serves as a repository for these models and provides a classification system predicated on diverse indicators, including likeness-related tags like 'Celebrity'. More recently, a Japan-based website called 'Petapi' (https://petapi.ai/) introduced a marketplace exclusively dedicated to the distribution of outputs generated by customised models, predominantly selling AI generated photorealistic and sexualised imagery of women. These examples illustrate the growing phenomenon of likeness generation and the online communities and practices that emerged around it. 

Image generation models fundamentally change the ability to manipulate likeness, but they do not introduce a new concept. Likeness manipulation predates the use of photo editing tools such as Photoshop \cite{Fineman}. The previous leap in this field materialised with the introduction of the AI-based deepfake production technology, which predominantly used GANs \cite{Goodfellow2014} to power face-swap systems that plausibly merge photorealistic images and videos of people into new renditions \cite{Westerlund2019}. The first recorded use of AI to manipulate likeness was performed by a Reddit user, using the username ``deepfake'', to swap the faces of celebrities into pornographic videos \cite{Hern}. Although outputs of diffusion models featuring an identifiable person’s likeness are often referred to as 'deepfakes' in both academic literature and media \cite{Ricker2023, Barr2023}, this paper contends that the advent of generative AI systems introduces a new era in visual likeness manipulation, due to fundamental conceptual and practical differences instituted by them. 

Diffusion-based generative models are trained to decode objects based on the association between their input content (pixels) and labels. Such approaches are conceptually different from previous technologies. Generative models they can distinguish, memorise and recreate a person’s likeness as a latent construct rather than merely warp a image representing the person. Where a deepfake is often meant to pass off the image as real, a likeness can be included in any manner of generative image. As such, likeness generation is not visually constrained to recognisable pre-existing content.

In addition to the technological differences, there are downstream differences in use cases precipitated by these models. First, generative models significantly reduce technical entry barriers. Any literate individual with internet access and limited source material can now generate a person's pre-encoded likeness. Conversely, previous technologies required considerable technical acumen to yield likeness depicting outputs in varied scenarios, and only removed barriers for the production of likeness in narrow use cases \cite{Leibowicz}. Secondly, the photorealistic output fidelity of generative models is hailed as unrivalled compared to previous technologies, rendering depictions of human subjects that are nearly indistinguishable from real ones \cite{nightingale2022ai}. Thirdly, the development and propagation of image generation models by leading technology companies have precipitated a surge in the distribution and popularity of such technologies. Fourthly, unlike their predecessors constrained to specific use cases or pre-existing content, generative models offer the potential to depict likeness in virtually countless contexts. Finally, the provenance of photorealistic outputs yielded by generative models prove more challenging to detect than those created using prior methodologies \cite{Ricker2023}. 

\section{Defining Likeness Through a Legal Lens}

Long before the emergence of generative AI, individuals sought to protect the use of their likeness or prevent others from using it in defamatory or exploitative manners. This rich history of likeness related legal disputes enables us to identify common properties in the legal interpretations of likeness. This section will undertake this inquiry of key cases related to the rights of an individual's image. It focuses on Western legal cases, as those were identified as key disputes that are cited across multiple jurisdictions. While it does not purport to be comprehensive of all legal systems, it seeks to provide a shared understanding of the limits of the term "likeness". 

The exploration of what constitutes an individual's likeness and the boundaries of its protection have been a legal quandary for well over a century, originating from a French court ruling in 1858 which ordered the seizure of unauthorised illustrations that depicted the likeness of a then-renowned actress on her deathbed \cite{Logeais}. The legal construct of likeness is typically embedded in "Personality Rights", a class of rights safeguarding against the exploitation of an individual's identifying characteristics and are recognised and upheld by the tort laws of numerous jurisdictions \cite{Resta}. 

Likeness is often invoked by two separate but related personality rights. The first is a right against a misappropriation of one's likeness for the protection of privacy rights \cite{Riley}, thus it narrowly prohibited to use a person’s likeness in manners that compromise the integrity of an individual’s identity \cite{Dogan}. The second, known as the right of publicity, expanded the initial privacy-rooted right into the protection of the property rights of public figures whose likeness carries a commercial interest, granting them control over the commercial exploitation of their likeness \cite{Lee-Richardson}. While the right of publicity is US-centric, where it is a non-federal right adopted by approximately half its states \cite{Petty}, the privacy-oriented right against misappropriation was acknowledged in a variety of other jurisdictions \cite{Brüggemeier}. Certain jurisdictions, such as the United Kingdom, do not officially recognise personality rights, but offer protection against likeness misuse under other legal instruments, such as the ‘passing-off' tort against the misrepresentative commercial use of one’s likeness \cite{Stallard}. 

In Midler v. Ford, the United States Court of Appeals clarified that likeness pertains exclusively to visual images and not to identifying attributes of individuals in other modalities such as their voice \cite{MidlerFord}. However, the ambiguous delineation of visual traits constituting an individual’s likeness has produced legal disputes over time. French legal theorists argued over half a century ago against recognising a person’s right to their likeness due to the complexities associated with proving that a picture represents a specific person \cite{Wagner}. 

In a more recent case, Brophy v. Almanzar \cite{Brophy}, the plaintiff contended that the American rapper Almanzar, known as "Cardi B", misappropriated his likeness over her album art cover by portraying his back tattoos on a model in a lewd pose with his face unseen. Like other landmark right of publicity cases in the United States, including Elvis. v. Capece \cite{Mays} and Midler, the court in the Cardi B Affair applied a subjective standard of "likelihood of confusion" between the depicted person and the plaintiff in order to adjudicate what constitutes likeness. Citing from Abdul-Jabbar v. General Motors, the court in the Cardi B Affair decided that adjudicating the likelihood of confusion is a subjective "question for the jury", based on whether the depicted individual was "sufficiently identifiable" with the plaintiff to create a potential for confusion. The Cardi B case, along with Midler, Elvis and Abdul-Jabar, highlights the absence of objective criteria for evaluating identity appropriation, necessitating reliance on subjective judgement steered by the likelihood of confusion.

The subjective nature of likeness appropriation was underscored in a series of cases where "look-alike" individuals resembling celebrities were allegedly misused by commercial companies to mislead audiences into believing they viewed the celebrity. In 2019, singer Ariana Grande sued clothing company Forever21 for featuring a model who allegedly looks "strikingly similar" to Grande, while wearing clothing and hairstyle similar to those worn by Grande in a music video \cite{Arianagrande}. Media personality Kim Kardashian lodged a lawsuit against clothing company Gap for featuring a "celebrity look-alike" in their advertising campaign \cite{KimK}. In the same year, the Israel Supreme Court adjudicated a case between Nespresso and its Israeli competitor, Espresso Club, who featured an actor resembling Nespresso's presenter, George Clooney, in a commercial. The court held that a reasonable viewer would discern no likeness between the two, as the intent was not to imitate but to ridicule Nespresso \cite{Nespresso}. It appears, then, that even when actual similar likeness exists, the potential to confuse coupled with an intent to do so created grounds for legal contention. 

The legal corpus, then, highlights the subjective nature of the interpretation of likeness. It is a consensus of viewers that defines likeness, rather than any single objective trait. Based on this key property, we define likeness as: 'a visual depiction that sufficiently identifies a person and is likely to create uncertainty among viewers as to whether it represents the identifiable individual'. The importance of this definition is that it confines the discussion in the following chapters to harms and considerations that relate to outputs that are likely to be perceived as an identifiable person. They do not address instances where certain visual traits of people were used, or even their entire likeness, in manners that is not likely to identify them. Such harms may still be addressed under other privacy considerations, but they remain out of scope for this paper.

\section{A Taxonomy of Likeness Generation Harms}

A fundamental gap in informing future sociotechnical literature is the identification of likeness generation harms. We address this by proposing a harms taxonomy that could be expanded in the future to include other AI generated media modalities, such as video or audio. The proposed taxonomy finds its inspiration in the burgeoning literature mapping the harms arising from generative language models \cite{Weidinger2021}, and was compiled by conducting a systematic meta-analysis of an established body of literature addressing online harms, predominantly those of deepfake technologies. 

We introduce a taxonomy of seven harm categories. Some categories introduce ancillary and derivative harms, as illustrated by Figure 1 below. The taxonomy is structured around the nature of the harm (e.g. reputational harm) rather than the means by which it is inflicted (e.g. impersonation). As some harm categories can conflate within certain use cases, categories were made discrete if they significantly diverged in their source, shared targets and audiences or use cases. For example, the generation and distribution of a nude likeness depiction can concurrently invoke sexual harms (e.g. sexual harassment) and reputational harms (e.g. defamation). Finally, the source of the harm was specified for each category. As noted below, most harms materialise only in the distribution of generated likeness and not in its very generation, as the awareness of the user generating likeness about its artificial provenance often diffuses its harmful implications.

\begin{figure*}[t]
    \centering
    \includegraphics[width=\textwidth]{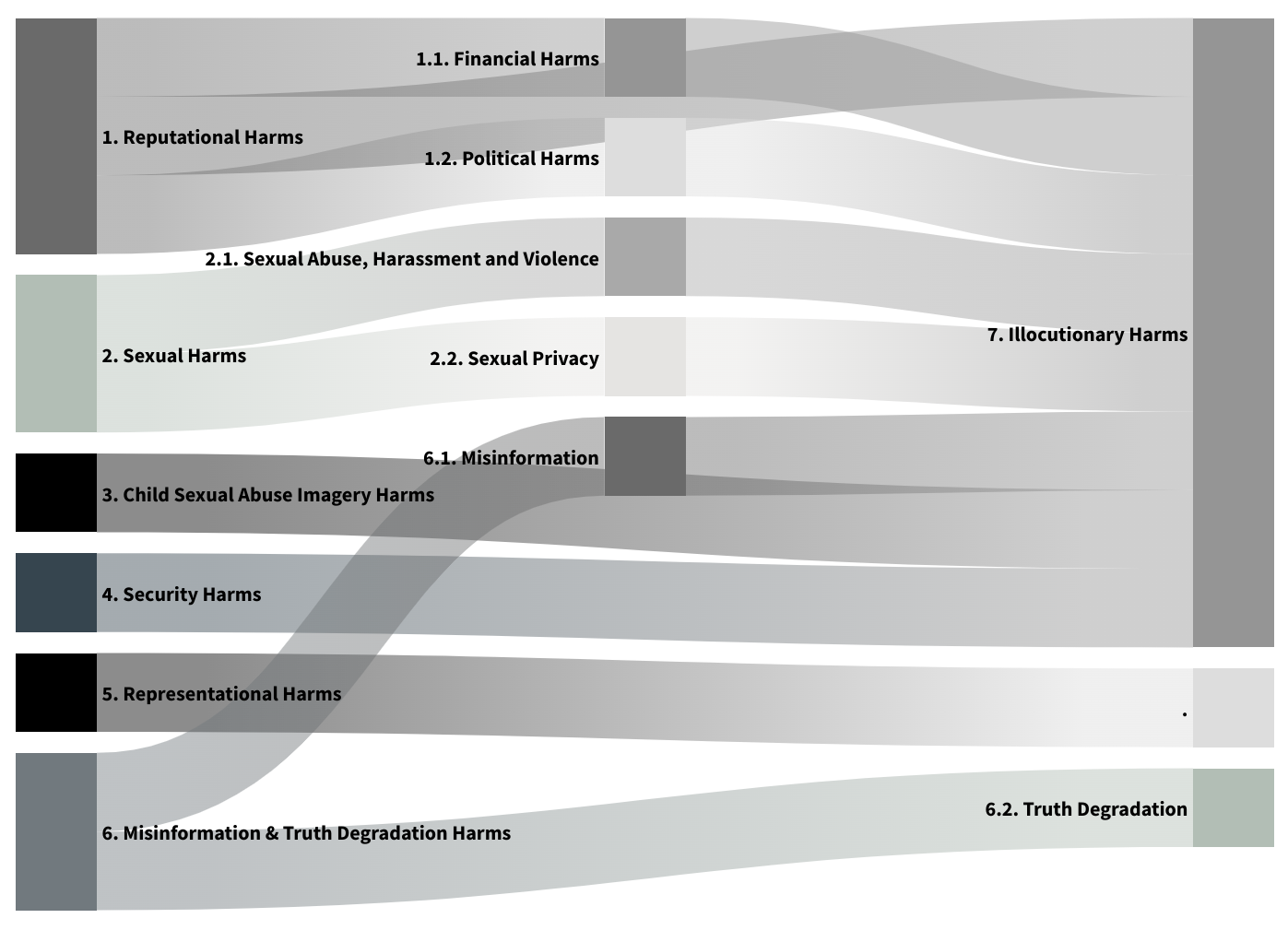}
    \caption{Visualisation of the Likeness Generation Harms Taxonomy.}
    \label{fig:enter-label}
\end{figure*}

\subsection{Reputational and Ancillary Harms}

Reputational harms occur when an individual's likeness is generated and disseminated in a manner that undermines their public standing. They arise when generated likeness is distributed in manners that defame, deface or otherwise tarnish the depicted individual’s character. The literature presents various use cases in which manipulating a person’s likeness damages their reputation. These often relate to public figures and involve the photorealistic depiction of likeness in sexually explicit settings \cite{Gieseke}, humorous and satirical contexts \cite{Adjer2019} or embarrassing, humiliating or otherwise disparaging scenarios. 

Emerging examples of reputational harm stemming from image generation models’ outputs include images of actor Will Smith portrayed as Adolf Hitler in a fictitious Netflix series \cite{Ibrahim2023b}, and Pope Francis meeting with "satanic priests", fueling conspiracy theories online \cite{Kasprak2023}. Image generation models are particularly prone of generating potentially reputation damaging outputs, even from benign prompts, due to a phenomenon known as ‘inappropriate degradation’, increasing their likelihood of producing inappropriate outputs based on their training on internet-scraped data, which is often biased towards degenerate human behaviour \cite{Schramowski2023}. 

Alongside the intrinsic negative effect of reputational harms, they may also introduce ancillary negative externalities rooted in the reputational property. Courts have recognised the potential of reputational harms to cause financial harms due to the economic value associated with an individual's public image \cite{Spivak2019}, particularly in the case of public figures, given their perceived higher economic value associated with their reputation \cite{Grady1994}. In the context of image generation models, likeness of public figures is more likely to be generateable due to the availability of their labelled images in public datasets. Alongside public figures, the advent of accessible sampling techniques like LoRAs also extends the risk of financial harm to private individuals. Previous research focusing on Deepfake technologies indicated that private individuals who had their likeness manipulated in sexual contexts experienced detrimental consequences in their employment and job-seeking endeavours \cite{Gieseke}.

Reputational harms can also indirectly invoke political harms, as seen in past instances where the likeness of politicians and other public figures was often manipulated in contexts that could influence political processes \cite{Diakopoulos2020}. One such example is the manipulated image of Burmese politician Aung San Suu Kyi donning a hijab amidst local tensions between Buddahist and Muslims \cite{AlteredImages2014}. Similar harms have already emerged in the context of image generation models. Examples include photorealistic depictions of Donald Trump being arrested in the backdrop of his criminal indictment \cite{DiPlacido2023}, and the Dalai Lama being arrested amidst controversy around his interaction with a child \cite{Ibrahim2023a}. Simultaneously, experts warned that the democratisation of image generation models is likely to affect the upcoming 2024 US elections due to their likeness generation capabilities \cite{Benson2023}.

\subsection{Sexual harms}

Image generation models also propagate harms that are intrinsic to the sexual nature of images sexually depicting a person’s likeness, manifesting with both their distribution and very generation. It is likely that likeness generation will be used to produce sexually explicit outputs, as deepfake technologies were previously used disproportionately for such purposes. A 2019 report estimated that 96\% of all Deepfake videos online involved non-consensual pornographic representations of individuals \cite{Adjer2019}. A separate report corroborated this trend, indicating that 90-95\% of Deepfake video outputs depicted non-consensual pornography, 90\% of which targeted women \cite{Hao2021}. 

Survivors of sexually explicit deepfakes have described their targeting as sexual abuse \cite{Martin2022}, sexual violence and sexual harassment \cite{Hao2021}. Sexual deepfakes have also been weaponised to silence activists campaigning against sexual violence \cite{Maddocks2020}, and to instigate threats of sexual violence towards public figures \cite{Gieseke}. While these harms are contingent on distribution, another harm in the form of infringements to sexual privacy stems from its generation. Chesney and Citron \cite{ChesneyCitron2018} describe this harm as the exploitation of a person’s sexual identity for another person’s gratification. Despite the outputs not depicting the real body of the subject, theorists argued that it infringes their privacy through the repurposing of their identity into a fictitious sexual one \cite{Gieseke, KuglerPace2021}.

Alongside purposeful likeness generation for sexual purposes, image generation models are likely to exacerbate sexual harms due to their reliance on sexualised datasets. Recent research concerning the Stable Diffusion model shows that it suffers from a sexual objectification bias towards women, being 42\% likely to retrieve a sexualised depiction of women compared to 9\% for men \cite{Wolfe2023}, which aligns with the aforementioned internet data driven phenomenon of ‘inappropriate degradation’ of image generation models. Scaling sexual objectification across society through such biases can indirectly exacerbate physical and mental harms rooted in sexual objectification such as shame and anxiety \cite{FredricksonRoberts1997}. 

\subsection{Child Sexual Abuse Imagery Harms}

Harms stemming from Child Sexual Abuse Imagery (CSAI) are outlined as a distinct category due to societal perceptions that the very generation of CSAI constitutes harms, rather than solely its distribution. Using legal systems as an indicator to societal norms, CSAI presents the only category of generated likeness harms that is criminalised in some jurisdictions for its private storage and viewing. Section 1 of the United Kingdom Protection of Children Act \cite{ChildrenAct1978}, for instance, criminalises the possession of indecent photos that convey the impression that the person depicted is a child, if it is possessed with a view to be distributed or shown, even if to the possessor by themselves.

With the proliferation of image generation models, security firms have reported instances of child predators manipulating existing images of children to create ``novel pseudo-photographic'' CSAI \cite{PaltieliFreud2023}. Similar findings were supported by a Washington Post investigative report on image generation models and CSAI \cite{Harwell2023}. The Australian eSafety commissioner similarly warned about the use of generative AI could be used by child predators, and that her office has already received reports about cyberbullying of children and image-based abuse \cite{Butler2023}. Finally, with the emergence of previous deepfake technologies, creators of pornographic Deepfakes expressed concern over the inclusion of children’s images in their ``faceset'' datasets of facial images \cite{Cole2018}. This concern could be exacerbated for image generation models capable of producing explicit content, which are trained on large volumes of data that is likely to contain depictions of children.

\subsection{Security Harms}

Security experts underscore the risks posed by likeness generation to personal security. Synthetic imagery featuring likeness in compromising contexts were specifically outlined as tools for extortion and blackmail \cite{Bateman2020}. Fraudulent activities have also employed likeness reproduction, using deepfakes featuring the likeness of celebrities to promote dubious products \cite{Kropotov2022}. 

With the release and growing use of image generation models, the United States FBI issued an alert against a rising use of malicious actors in benign photographs of people available online in order to harass or ``sextort'' them, cautioning the public against uploading their photo online \cite{FBI2023}. The wide availability of image generation models and the lack of technical acumen needed for the production of outputs also positions children as a group susceptible to its security harms, as it provides the means to conduct cyberbullying through sampling images of children or the simpler inpainting functionality. 

\subsection{Representational Harms}

A harm unique to image generation models compared to previous likeness manipulation technologies is its inadvertent propensity of causing representational harms, being the generation of images depicting a person’s likeness in manners that underrepresent the context of their culture or identity. The literature has long shown the propensity of algorithms to exacerbate inequality \cite{FriedmanNissenbaum1996}, which in the context of artificial intelligence propagates pre-existing social biases \cite{Wachter2021}. Such biases have already been demonstrated in image generation systems, as the literature shows that they amplify demographic stereotypes, misrepresent prompts related to certain countries or make cultural assumptions that manifest in the lack of diversity of generated outputs \cite{Bianchi2023}. 

In the context of likeness generation, image generation models can exacerbate bias in various manners. Primarily, they could contribute to an over or under representation of likenesses of people from different geographical regions, based on the availability and inclusion of relevant datasets. They can also misrepresent certain features in such likenesses based on the cultural context of the training data. For example, if trained on images of people in cultures where the norm is to smile, they may globally portray images of people smiling. Image generation models have been shown to associate certain professions with men and vice versa \cite{Cho2023}, such biases may also be carried over in depictions of specific humans. They may also misrepresent physical characteristics of a person in manners that are more prevalent in training data, such as stereotypically white ideals of attractiveness \cite{Bianchi2023}. Models may also associate the generated likeness with other objects that do not represent their culture, but were abundant in the training data. Consider, for example, prompting for the generation of Japanese Emperor Naruhito eating lunch. Models may fail to capture the local concepts of life in Japan based on its training data, and thus lead to the misrepresentation of people.

\subsection{Misinformation and Truth Degradation Harms}

A fundamental element associated with the distribution of generated images not disclosing their provenance is their misinforming potential. As highlighted in the preceding discussion on reputational harms, images featuring one's likeness have been disseminated in ways that politically mislead viewers. Misleading visual manipulations of likeness produced by previous deepfake technologies have been exploited by a wide range of actors including online trolls, bots, conspiracy theorists, hyper-partisan media and foreign governments to influence political, geopolitical, commercial or sensationalist interests \cite{Masood2023}. 

Likeness generation may also contribute to the erosion of truth in online spaces. Users have already been shown to growingly question the authenticity of the imagery they view, a phenomenon that has been previously termed as ``reality apathy'' \cite{Westerlund2019}. Distribution of generated likeness may play a significant role in blurring the authenticity of imagery online, given its introduction of mass likeness generation potential and its aforementioned property of its outputs being harder to detect.

\subsection{Illocutionary Harms}

Generated likeness harms are also likely to cause illocutionary harms stemming from many of the aforementioned direct harms. Drawing from J.L Austin’s linguistics doctrine, Rini and Cohen \cite{RiniCohen2022} define illocutionary harms as the act of illegitimately compelling a person to perform an undesired speech act to deny the veracity of an allegation. This, according to Rini and Cohen, manifests as public figures are forced to address fictitious imagery featuring their likeness, thereby undermining their communicative agency. While Rini and Cohen attribute illocutionary harms to public figures, illocutionary harms also affect private individuals despite their lack of access to public forums. Instead, private individuals are often forced to alter their communication by pursuing online platforms for the removal of harmful depictions of their likeness. This is often a slow and inefficient process due to the safe harbour regime governing platforms, which places the heavy burden of refutability on the shoulders of the victim and have been shown to be particularly sluggish in the context of deepfake-related enforcement \cite{O’Donnell2021}. These processes are likely to adversely affect more people as the availability of likeness generation scales. 

Illocutionary harms may also manifest at a societal, non-individual level. The very introduction of generative systems capable of reproducing one’s likeness creates a chilling effect for people’s online participation. This was demonstrated by a recent commercial published by the German Deutsche Telekom telecommunications company \cite{Houston2023}, warning users against uploading their likeness to social media in light of AI capabilities. Such concerns create a ``negative'' illocutionary harm of avoidance, where a person might have wanted to participate.  

\section{Considerations for Mitigations Deployment}

The categories of harms identified, when considered alongside the increasing capabilities and distribution of generative models, underscore the need to develop new mitigation pathways. However, those should also be carefully designed to enable beneficial or legitimate use cases of this emergent capability.

Mitigation deployment for likeness generation is challenged by the multitude of actors that are involved in potential likeness generation harms. Taking analogy from the recent proposal of the generative AI supply chain as a conceptual apparatus \cite{lee2023talkin}, likeness generation can be seen as a multilayered multilayered system involving occasionally distinct actors with upstream and downstream roles involved in producing a potentially harmful synthetic image featuring a likeness. At the likeness generation layer the supply chain includes data curators, model developers, system deployers and third party developers that further fine-tune or retrain models. At the distribution layer, actors include users, model sharing platforms and output distribution platforms (e.g. social media). 

In pursuit of clarifying the appropriate mitigations and their deployment, we identify and surface four criteria for mitigations efforts that stem from the common features of the harm categories identified. We view these criteria as parameters that should be weighted together and may point at a similar intervention junction, or surface various equally important touch points for mitigating the harm. In the latter scenario, it may be justified for multiple entities in the supply chain to address the same harm. 

\subsection{1. Indexical Sufficiency}

The first question that we propose asking is whether the model was capable of producing a likeness based on its training data, a phenomenon that we term ``indexical sufficiency''. This term refers to the capacity to non-arbitrarily generate a likeness indicating prior knowledge of the particular person. In most cases this would mean ``can you prompt for that specific person's likeness''. There is at least one point in the likeness generation process before which model was not indexically sufficient to generate likeness. That point will vary between different cases and model capabilities, as the cases below illustrate. For all cases, consider an output which includes a nude image of a person who can be identified, either based on their facial features or physical appearance.

\begin{itemize}
    \item The first case is text-to-image in a closed system. It involves prompting an image generation system for an image of a known person and of nudity and successfully obtaining a likeness. If this can be done without additional information about the subject, then the system has indexical sufficiency for both nudity and the specific subject. It thereafter uses the techniques of generative imagery to combine these in a plausible, potentially photorealistic, manner. Such systems typically contain closed source models and use malicious prompt detection to mitigate generation. However, such systems could also make use of open source models but simply prohibit further fine tuning or img2img capabilities. 
    \item The second case is text-to-image in an open system. These are typically systems with open source image generation models. These similarly train on vast often publicly available data but it they may not be sufficient to generate a specific person, much less them nude. However, open systems are extensible with additional models such as LoRAs, which themselves can be combined with base models to produce a likeness. Prior to the inclusion of the LoRA, the model was not indexically sufficient. However, the model inherently contains the potential to become indexically sufficient through extensible tools to train downstream models. While typically open source, one could also envisage a system with a closed source model that still allows fine-tuning via APIs.
    \item The third case is an image-to-image system. This is a system where the model does not need to be indexically sufficient for the likeness since the likeness is contained in a prior image. This would be the kind of system used in `nudify' apps where a reference photo is digitally manipulated, but the likeness is never indexed. In such cases, the system would have to have prior knowledge of nudity and body type. The likeness, by contrast, is established through the reference photo. Typically, these approaches cannot create flexible representations of the likeness, only alter areas of the image around the likeness to frame the subject in a different (and in this example, unwanted) manner.
\end{itemize}

In all cases the consequences are similar: there exists a representation post-process that harms a subject. Yet, in the three cases, the system that produced the offending image did so through different capacities and with a different level of flexibility.

Hence, in deploying mitigations for likeness generation, the first consideration that should be considered is the junction in which the likeness was introduced. For example, it may be warranted for actors at the model, data and system level to evaluate and mitigate for text-to-image harms. However, for harms involving third party manipulation (such as in Cases two and three), it may be warranted to mitigate the uses of distribution platforms or third parties. As was previously raised in the context of deepfake creation, there are challenges in applying controls downstream the supply chain \cite{widder2022limits}. However, future research clearly asserting the different responsibilities in the supply chain based on the indexical sufficiency criterion may assist in enabling nuanced mitigations that do not overblock legitimate use cases.

\subsection{2. Harmful Generation or Distribution}

Another criterion emerging from the harms taxonomy is a difference in the source of the harm. While some harms manifest with the very generation of a person's likeness, others only occur with the distribution of such likeness featuring outputs.

For example, a user generating a misleading image featuring the likeness of a politician may not introduce harm when done in the user's own private domain. However, such an output could introduce severe harms to democratic processes when distributed. Besides such misinformation harms, reputation and security harms are examples of harms likely to be rooted in the distribution of generated likeness. Other harms, such as some sexual harms, child safety and representational harms may occur with the very exposure of the generative model user to them.

We view this distinction as another key criterion in nuanced mitigations development and deployment. Harms occurring at the generation level can only be prevented by mitigations deployed by more upstream actors in the supply chain, such as the data curation, model development, and system deployment processes. For harms anchored in the distribution of the output, assigning mitigations responsibility to post-generation supply chain actors, such as social media distribution platforms, or the end users themselves, could enable the effective mitigation of likeness generation harms.

\subsection{3. Denoted or Contextual Harms}

The third criterion relates to whether the nature of the harm is direct or implicit. Some images require additional context in order to be understood as harmful. 

The philosopher and semiotician Ronald Barthes previously highlighted that signs, such as imagery both denote a fact and connotes an impression  \cite{Barthes}. Denotation constitutes a sign's primary, literal meaning, and connotation represents an implicit, secondary, and value-laden meaning. We do not only care about what we see (denotation) but also about what’s implied (connotation). 

Some harms do not require their observer to have additional context in order to infer a harmful connotation. For example, nude elements in an image could be harmful to the depicted person if viewed by anyone, regardless of their contextual understanding. However, other outputs require context in consider to establish a harm. Consider the aforementioned example of the output depicting police officers arresting Donald Trump. Without sufficient context about Donald Trump or his ongoing trial, the potential harm to democratic processes may not materialise. It is also here where we remind that the harms from generative imagery are not simply about whether the image is deemed photorealistic-but-fake, but about what the image connotes relative to the subject.

We believe that outputs that connote contextual harms require a federated responsibility approach across the supply chain. Firstly, contextual harms often connote harms that can be mitigated to a large extent by identifying their synthetic nature. In the aforementioned depiction of Donald Trump, sufficient watermarking could indicate that the image does not in fact represent a real event. As such, the generation layer could have a responsibility to watermark such outputs. The distribution layer, in turn, could surface its synthetic nature based on the watermarking to diffuse the harm. Legal scholars similarly suggested to tackle online misinformation with transparency mechanisms rather than content adjudication \cite{WoodRabel2018}. Secondly, proactively predicting contextual harms that have no clear visual indications, may pose challenges for upstream mitigations. The distribution layer could exercise a contextual analysis in adjudicating the content. However, nascent techniques can be developed to enable such contextual filtering upstream. For example, by utilising other advancement in generative AI such as image understanding by vision language models to produce textual context about the generated image, as an input for upstream filters.

\subsection{4. Benign or Harmful Prompts}

Finally, another mitigations consideration that we will mention briefly is the nature of the prompt. We argue that the responsibility of likeness featuring outputs changes based on whether the prompt was benign or harmful.

Harmful prompts generating harmful content involve an an agent that had intent of generating such an output. Exposing the user to such harmful outputs, assuming that they are not further distributed or introduce harms to a third person, might affect the harmfulness of the output. In cases where the user is malicious and intends to cause harm with such output, they also share responsibility in the harmful likeness generation. Furthermore, for such harmful prompts, downstream mitigations such as input filters  may be efficient \cite{Rando2022}.

However, there could be cases where benign prompts produce harmful likeness. Consider a theoretical scenario where the prompt 'person' generates a defamatory output featuring a known person's likeness. For those cases, mitigations are best suited upstream. Such mitigation could include more careful data filtering or disclaimers to system developers about the safety performance of the model on transparency mechanisms such as model cards \cite{Mitchell2019}.

\section{Future Research Directions and Limitations}

This paper introduces an initial exploration into definitions, harms and policy questions related to likeness generation. The societal implications of likeness generation is a nascent research area which has ample theoretical and empirical gaps that require additional work. Future research may examine how likeness generation and distribution is used in communities of practice, and whether harms actually manifest online. It may also address the issue of likeness related evaluations. While proposed frameworks for multimodal generative AI safety evaluations begin to emerge \cite{Weidinger2023}, they do not propose evaluating model capabilities to generate likenesses. Methodologies for measuring and evaluating models for their ability to generate likenesses of different publicity levels could assist in improving the safety of generative models. Thirdly, additional work is required on the societal implications of exisitng mitigations. Some mitigations discussed in this paper may not be socially desirable due to various ethical and legal considerations, or conversely, may be warranted but face implementation challenges. For example, detecting likenesses would require the processing of biometric data or other source material even if done in a privacy aware manner through technique such as hashing. Finally, while this paper focused on likeness generation in imagery, the increasingly multimodal nature of generative AI models will require adapting this work to likeness generation in additional modalities, notably audio and video.  

The limitations of the ideas proposed in this paper should also be acknowledged. The proposed harms taxonomy is inductive, and does not capture all possible harms and their common characteristics. It aims to group harms in a systematic manner that would serve future work requiring a consolidated analysis of generated likeness harms. The taxonomy is also anchored in a definition that was surfaced by focusing on specific legal systems, and may not fully cover legal and philosophical concepts of likeness in other jurisdictions. It should also be acknowledged that some of these harms are inferred based on previous trends such as deepfakes, that may not equally manifest in the context widespread and easy-to-use image generation models.

\section{Conclusion}

In this paper, we conducted an inquiry into the novel phenomenon of generated likeness. We addressed likeness generation as a sociotechnical process, highlighting its technical properties, its multilayered supply chain, and how it differs from existing likeness manipulation technologies. We referred to the legal literature in search for an operational definition of likeness. 

Against the backdrop of these technical and legal literature reviews, a taxonomy of likeness related harms was proposed, which was surfaced through a meta analysis of relevant online harms literature, and sorted by a systematic identification of harms commonalities. The proposed taxonomy is not a deductive classification of all possible harms, but a way to delineate and aggregate potential harms based on common attributes and saliency. The taxonomy outlines seven categories of harms: reputational, sexual, CSAI, representational, misinformation and truth degradation, and illocutionary. It supplemented the theoretical harms with real-world examples manifesting in similar technologies, alongside the early emergence of such harms by generative AI systems. 

From this work we attempt to address a gap in developing and deploying mitigations for likeness generation harms. Absent of clear criteria that should be weighed in deploying harms, it is unclear which actors or mitigations should focus on addressing the harms outlined at the core of this paper. Based on the properties of the likeness generation and its harms, we propose four such criteria that could help operationalise mitigation development. Those include the concept of indexical sufficiency; a distinction between likeness distribution and generation; differentiating denoted harms from contextual harms, and; considering whether the prompt was benign or harmful. 

We believe that the future use of a subject’s likeness in a generative manner will become an increasingly common practice both within creative industries, where due consideration of the likeness was a central part of the 2023 American Writer’s strike \cite{Dalton2023}, among private citizens for purposes ranging from entertainment to editing of personal photos, to various other entities. Technologies like Stable Diffusion and Midjourney provide a foreshadowing of the plasticity of the likeness through generative AI across many modalities, but also of the practical challenges in managing such harms given the complexity of the likeness ``supply chains'' and where indexical sufficiency resides for different subjects. Thus, identifying the harms, their characteristics, and their various other criteria outlined in this paper will enable the industry, users, and the public to help steer best practices and regulatory focus with the ambition of allowing people effective and creative representation instead of fears of misrepresentations.

\bibliography{aaai24}

\end{document}